\documentclass[journal]{IEEEtran}
\usepackage[utf8]{inputenc}
\usepackage{cite,amsmath,amssymb}
\usepackage{graphicx,epstopdf}
\usepackage{epsfig}
\usepackage{textcomp}
\usepackage{xcolor}
\usepackage{hyperref}
\usepackage{cleveref}
\usepackage{orcidlink}
\usepackage{stfloats}
\usepackage{caption}
\usepackage{adjustbox}
\usepackage{subcaption}
\usepackage{lipsum}
\usepackage{enumitem}
\allowdisplaybreaks

\begin{document}

\title{AI Enabled 6G for Semantic Metaverse: Prospects, Challenges and Solutions for Future Wireless VR}

\author{%
Muhammad~Ahmed~Mohsin{~\orcidlink{0009-0005-2766-0345
}},
Sagnik~Bhattacharya{~\orcidlink{0000-0003-0385-1225}},~Abhiram~R~Gorle{~\orcidlink{0009-0006-3028-1626}},~Muhammad Ali Jamshed{~\orcidlink{0000-0003-0385-1225}},~and~John~M.~Cioffi{~\orcidlink{0000-0003-1353-8101}
}%

\thanks{Muhammad Ahmed Mohsin, Sagnik Bhattacharya, Abhiram R. Gorle and John M. Cioffi are from Stanford University. Corresponding author: \{muahmed@stanford.edu\}.
Muhammad Ali Jamshed is with the College of Science and Engineering, University	of Glasgow, UK (e-mail: muhammadali.jamshed@glasgow.ac.uk).}.}

\maketitle

\begin{abstract}
Wireless support of virtual reality (VR) has challenges when a network has multiple users, particularly for 3D VR gaming, digital AI avatars, and remote team collaboration. This work addresses these challenges through investigation of the low-rank channels that inevitably occur when there are more active users than there are degrees of spatial freedom, effectively often the number of antennas. The presented approach uses optimal nonlinear transceivers, equivalently generalized decision-feedback or successive cancellation for uplink and superposition or dirty-paper precoders for downlink.  Additionally, a powerful optimization approach for the users' energy allocation and decoding order appears to provide large improvements over existing methods, effectively nearing theoretical optima. As the latter optimization methods pose real-time challenges, approximations using deep reinforcement learning (DRL) are used to approximate best performance with much lower (5x at least) complexity. Experimental results show significantly larger sum rates and very large power savings to attain the data rates found necessary to support VR. Experimental results show the proposed algorithm outperforms current industry standards like orthogonal multiple access (OMA), non-orthogonal multiple access (NOMA), as well as the highly researched methods in multi-carrier NOMA (MC-NOMA), enhancing sum data rate by 39\%, 28\%, and 16\%, respectively, at a given power level. For the same data rate, it achieves power savings of 75\%, 45\%, and 40\%, making it ideal for VR applications. Additionally, a near-optimal deep reinforcement learning (DRL)-based resource allocation framework for real-time use by being 5x faster and reaching 83\% of the global optimum is introduced. 

\end{abstract}


\section{Introduction}
\label{sc:introduction}
\begin{figure*}[t!]
    \centering
    \includegraphics[width=\textwidth]{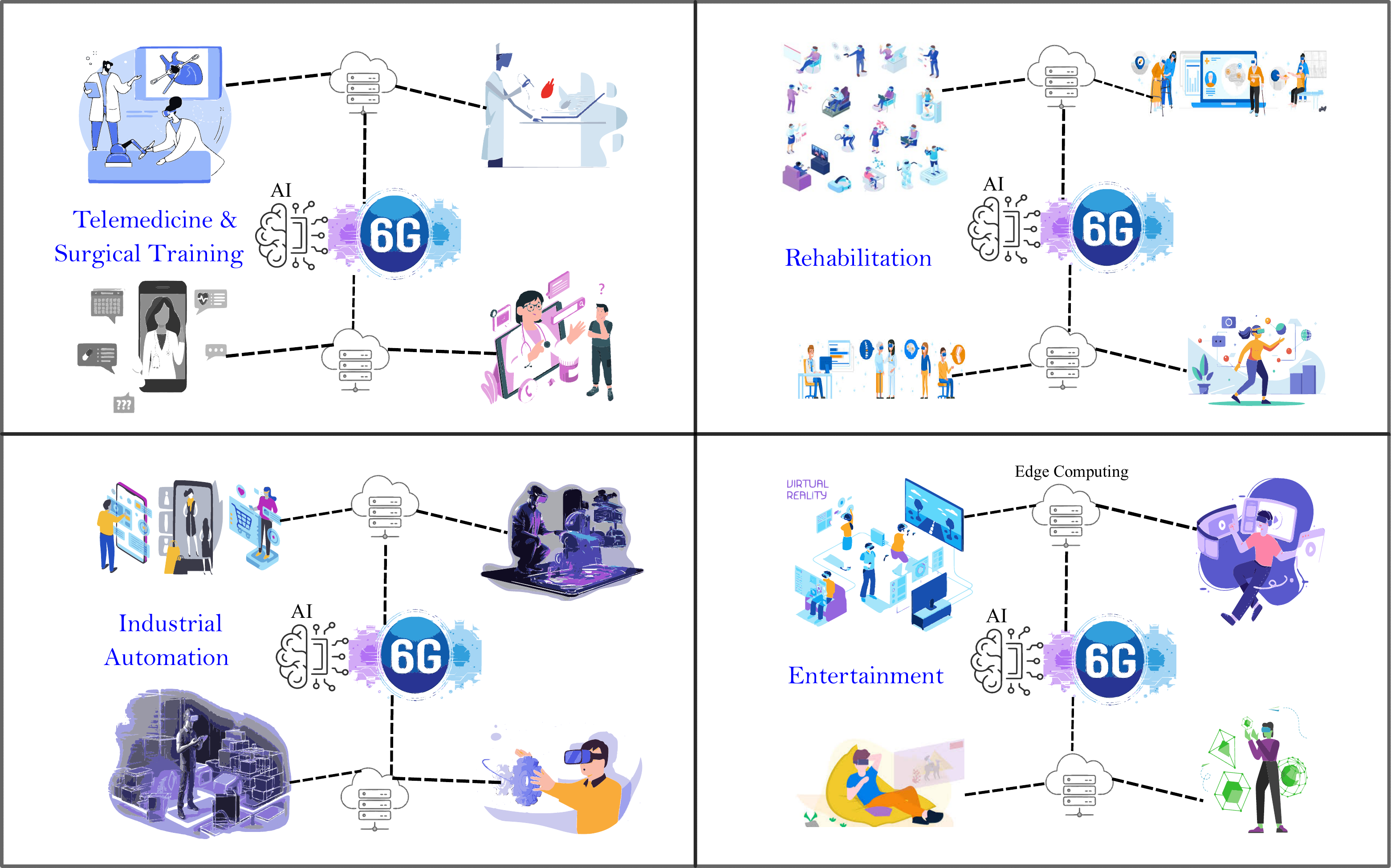}
    \caption{Wireless VR Use Cases.}
    \label{fig:wireless_VR}
\end{figure*}

The Metaverse, a term coined by Neal Stephenson in his science fiction novel~\cite{allouche2024technology}, has evolved into a foundational technology that represents real-world systems digitally, enabling interactions with avatars, smart homes, and immersive technologies. Metaverse, being a vast term, encompasses technologies like virtual reality (VR), augmented reality (AR), blockchain, brain-human interfaces, etc., and has several important applications such as telemedicine, entertainment, and industrial automation as shown in Fig.~\ref{fig:wireless_VR}. The Metaverse ecosystem has grown into an \$80 billion industry according to Goldman Sachs, comparable to the PC market~\cite{8038375}. As an example, Apple's recent Vision Pro wireless VR product enables wireless VR use cases with projected annual sales of \$1.7 billion. As a key metaverse component, VR differs from AR in that VR creates fully immersive computer-generated environments, while AR overlays digital information onto an actual environment. While traditional VR systems, like Pimax Crystal, are limited by wireline connectivity, wireless VR emerges as a promising solution and offers mobility support, processing-offload capabilities, and user-friendly form factors~\cite{9363323}. Through wireless-VR head-mounted displays (HMDs), users can now navigate and interact within immersive virtual environments without mobility constraints, transforming how humans engage with their surroundings and connect across geographical boundaries. This paper proposes solutions and suggests challenges to support the Metaverse wirelessly (through cellular enterprise Wi-Fi).

Wireless VR advances pose significant realization challenges. Wireless VR video streaming necessitates the transmission of extensive 360-degree visual data while maintaining very low latency. Wireless VR quality of experience (QoE) requirements include a refresh rate of $\leq$ 60 frames per second (FPS) and a latency budget of 14 ms~\cite{9097455}. For wireless VR systems that provide seamless service over unstable connectivity, handover necessarily must handle uneven and interconnected traffic between uplink and downlink channels, and minimum real-time delivery of VR content services. Moreover, VR video streaming uses a projection technique that maps pixels from a surrounding viewing sphere onto a two-dimensional viewport, a process known as viewport rendering~\cite{9430902}. Viewport rendering requires a huge number of matrix multiplications and thus consumes more power and degrades battery health. Wi-Fi (or cellular enterprise) 6G Uplink transmissions need low-power resource-allocation solutions that meet the required data rate. Semantically compressed communications for AR/VR distributed systems that prevent exhaustive usage of bandwidth are also important.

\begin{figure*}[t!]
    \centering
    \includegraphics[width=\linewidth]{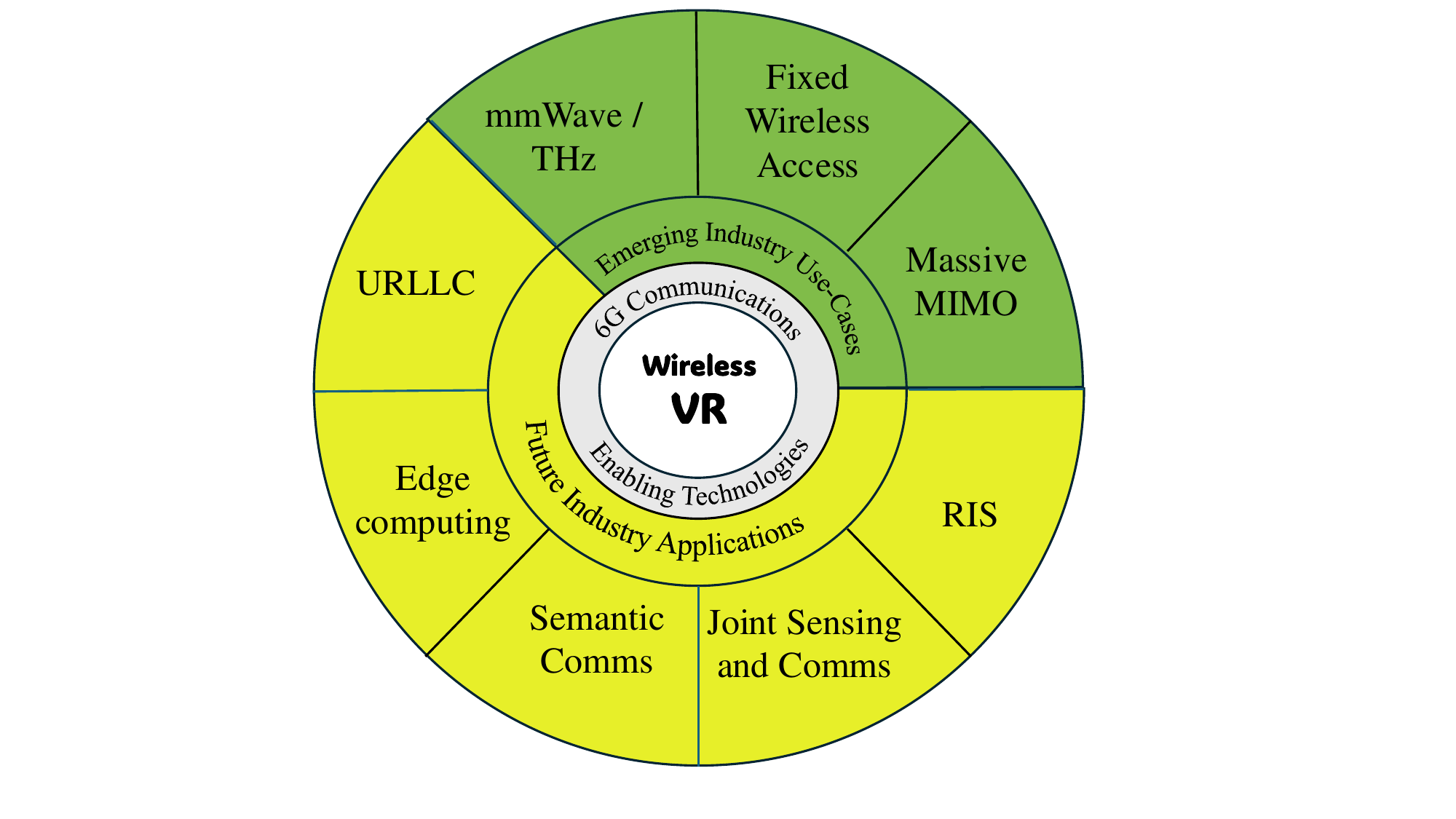}
    \caption{Wireless VR enabling technologies in 6G networks.}
    \label{fig:6G_technologies}
\end{figure*}

As the number of wireless VR subscribers grows, future wireless networks will face a low-rank channel-congestion problem. In the low-rank channel-congestion scenario, the number of active users exceeds the number of access point (AP) antennas (either Wi-Fi or cellular enterprise). In this paper, the term Beyond 5G (B5G) and 6G refers specifically to Wi-Fi and cellular enterprise networks within this context. This distinction is made because AR/VR systems are predominantly utilized in indoor Wi-Fi environments or through cellular enterprise networks. Additionally, our case study focuses on Wi-Fi and cellular enterprise networks as representative wireless environments, serving as the foundation for our experimental motivation. The current linear orthogonal multiple access (OMA) receivers used in industry standards, as well as the non-orthogonal multiple access (NOMA) methods, do not achieve the required sum data rates for these low-rank channel-congestion scenarios, and thus fail to meet minimum quality of service (QoS) requirements for wireless VR. B5G and 6G systems have enabled low-latency, high-data rate, low-power consumption, and ultra-reliable connectivity. 6G enabling technologies, as in Fig.~\ref{fig:6G_technologies}, like reconfigurable intelligent surfaces (RIS), integrated sensing and communication (ISAC), and semantic communications provide high data rate (500+ Mbps under Wi-Fi scenarios) that can enable VR applications. Furthermore, mobile-edge computing (MEC) for data offloading~\cite{8038375}, and caching~\cite{8003452} have been proposed as a viable solution to address wireless VR computing requirements. However, none of these technologies completely address the importance of receiver designs in achieving low-latency, high-data-rate solutions under low-power constraints. Optimization algorithms for resource allocation that minimize power transmission, performing effectively with rank-deficient channels, are necessary for wireless VR.
The article's contributions are:
\begin{itemize}
  \item A detailed overview of wireless-VR QoE requirements appears for indoor Wi-Fi scenarios, followed by a discussion of the challenges in meeting these requirements from a wireless perspective. Efficient solutions are then explored, culminating in a case study that demonstrates practical approaches for achieving wireless VR requirements.
  \item The case study proposes a wireless energy sum-minimization strategy for MAC/broadcast in congested AR/VR scenarios using Wi-Fi and cellular enterprise networks, which achieves high data rates at low SNRs by optimizing decoding order and time-sharing.
  \item An equivalent DRL based near-optimal resource allocation method for faster convergence in Wi-Fi scenarios is also presented to address real time deployment concerns.
  \item The optimum power sub-carrier allocation algorithm achieves 39\%, 28\%, and 16\% higher data rates than OMA, NOMA, and MC-NOMA while using 75\%, 45\%, and 40\% less power. 
  \item The DRL-based near-optimal resource allocation reaches 83\% of the global optimum 5x faster. 
\end{itemize}

\section{QoE Requirements for wireless Virtual Reality User} In Wi-Fi based virtual reality systems, user sensory experiences can be assessed through display clarity, visual scope, frame update frequency, and response time. These elements influence the performance criteria necessary to deliver an engrossing digital environment across diverse scenarios and a range of participants. This section discusses the QoE requirements from a Wi-Fi based wireless communications perspective.
\subsection{\textbf{Resolution Requirements}}
In wireless VR systems, Field of View (FOV) and resolution present significant challenges with limited network bandwidth. Wireless VR headsets typically support up to $150^\circ$ FOV, divided into central ($60^\circ$), peripheral ($30^\circ$), and monocular zones. Wider FOV demands higher resolution, increasing the strain on the wireless link to transmit large volumes of high-quality visual data at a higher data rate with minimal latency. The inability of current OMA and heuristic NOMA methods to satisfy these requirements often leads to compromises in resolution, device weight, and cost. 

Resolution, measured in pixels per degree (PPD), directly impacts the visual experience. Although the human eye requires 720 PPD for visual fidelity, current wireless VR headsets achieve only 20-30 PPD due to bandwidth limitations. Wider FoV further reduces PPD, exacerbating the “screen door effect,” where gaps between pixels become visible. Techniques like foveated rendering, which prioritize high detail in the central vision, partially alleviate these issues by reducing the data transmission requirements. However, achieving seamless FoV and resolution remains a challenge, especially in wireless networks constrained by limited bandwidth and latency concerns.

\subsection{\textbf{Latency Budget Requirements}}
In wireless VR systems, latency and refresh rate are crucial factors that significantly impact user experience. Latency, especially motion-to-photon latency—the delay between a user's movement and the corresponding visual update—should be $\leq$ 14 ms to prevent motion sickness and ensure seamless interaction. Latency comprises sensory latency (0.5–2 ms), input processing (1–3 ms), rendering (2–10 ms), and wireless frame transport (5–10 ms or more).

Refresh rate is directly linked to latency, determining how frequently frames are displayed. A minimum of 90 Hz is required to avoid motion blur and visual artifacts like screen tearing, with optimal experiences often needing 120 Hz or higher. In high-motion scenarios (e.g., 30°/s with high-resolution displays at 60 PPD), refresh rates up to 1800 Hz may be necessary, imposing substantial bandwidth demands on wireless networks. While higher refresh rates decrease motion-to-photon latency, they also escalate data transmission requirements.

To address these challenges, techniques such as asynchronous reprojection (interpolating frames for smoother updates), time warping (adjusting rendered frames based on head position), and low-persistence displays (reducing motion blur by minimizing pixel illumination time) are utilized. Nevertheless, achieving low latency in wireless VR remains a significant challenge, particularly due to limited bandwidth and low uplink transmission power.

\subsection{\textbf{Power Requirements}}For wireless VR, total power must remain below 1 W. AI compute power is a primary bottleneck, but cloud offloading offers a solution by transferring compute tasks to remote servers. While this reduces on-device computation to nearly zero, it increases wireless communication power consumption. If this communication power consumption can be minimized, cloud offloading becomes viable, allowing efficient VR operations within the power constraint and increasing battery life for AR/VR headsets. AR/VR headsets on average consume about 40\% of total compute and memory energy through projective transformation operations, and display, storage, and network components contribute roughly 20\% combined to total power consumption. WI-Fi-based cloud computing platforms can reduce computing powers at the cost of latency. Current OMA standards do not meet minimum data rate requirements for AR/VR at lower energy levels, contributing significantly to the overall power consumption. Solutions need efficient resource allocation strategies to address energy issues.

\section{Challenges for Wireless Virtual Reality: A Future Wireless Perspective}
Section II outlined the requirements for wireless VR in Wi-Fi scenarios. This section addresses challenges associated with meeting those requirements.

\subsection{Network requirements}
\subsubsection{\textbf{Data rates}}
For a seamless wireless VR experience, achieving high data rates is essential, especially for video streaming. Human perception standards require a minimum of 60 PPD resolution and a 120 Hz screen refresh rate. Transmitting 360° VR content with a 20:1 compression ratio necessitates a downlink speed of approximately 50.9 Gbps. Increasing compression can reduce the data rate but increases VR interaction delays due to added processing time. Furthermore, an uplink speed of 500 Mbps~\cite{bhattacharya2024optimalpowerallocationtime} or more is required to send motion data for VR, yet this must be achieved with low transmit power to preserve battery life in wireless VR devices. These demands are likely to become more stringent with higher resolution and FoV requirements, highlighting the need for advanced 6G measures to support wireless VR.

\subsubsection{\textbf{Communication latency}} In wireless VR systems, communication latency significantly impacts the quality of user experience. This latency includes communication delays, rendering delays, coding/decoding delays, and offloading delays. For a seamless experience and to avoid motion sickness, the previously mentioned total latency must be $\leq$ 14 ms. Huawei's study using a VR testing apparatus found that network transmission delay alone contributed 17.89 ms to the overall 82 ms lag time, meaning that wireless communication pathways are allotted only one-third of the target response time, adding considerable pressure on wireless systems to minimize latency~\cite{HuaweiCloudVR2024}.

\subsubsection{\textbf{AI Compute Latency}} Communication latency is directly linked to AI compute latency, as high communication latency makes offloading impractical. This requires AI computations to run locally on VR devices, creating a major latency bottleneck. However, if communication latency can be minimized, AI compute tasks could be offloaded to cloud servers, significantly reducing end-to-end latency and enhancing VR performance.

\begin{table*}[t!]
	\caption{Wireless Virtual Reality Headset Comparisons -- Visible Field of View (FoV), Resolution, Refresh Rate, IPD Range}
	\centering
	\begin{tabular}{|c|c|c|c|c|}
		\hline
		Virtual Reality Gear & Visible FoV & Resolution & interpupillary distance (IPD) Range (mm) & Refresh rate (Hz) \\ \hline \hline
		Meta Quest 3S & 97°H  -- 93°V  & 1832x1920 & 58--68 & 120   \\
		HTC Vive Focus Vision & 116°H -- 97°V  & 2448x2448 & 57--72 & 90  \\
		Sony SRH-S1 & /  & 3552x2840 & 60--75& 90   \\
		Roscosmos XR-2 & 157°D& 2880x2880 & 60--80 & 120  \\
        TCL NXTWEAR V & 108°D & 2280x2280 & 55--71 & 90  \\
	Apple Vision Pro	& 100°H -- 92°V & 3660x3200 & 51--78 & 100  \\
 Meta Quest 3	& 110°H -- 96°V& 2064x2208& 58--71 & 120  \\
PlayStation VR2	& 110°D & 2000x2040 & 58--73 & 120  \\
Somnium VR1	& 125°H -- 100°V & 2880x2880 & 60--76 & 120  \\
Samsung Odyssey	+	& 102°H -- 105°V & 1440x1600 & 90--72 & 90 \\
 \hline
	\end{tabular}
	\label{tab:icdar19}
\end{table*}
\subsubsection{\textbf{Error Probability}} Wireless VR headsets handle two primary categories of data: user movement signals and immersive content streams. The movement signals must be transmitted nearly error-free. Since the virtual environment is rendered in real-time based on the user's current position and actions, any inaccuracies in movement data could necessitate additional computational time to correct faulty pixels. Consequently, it is crucial to maintain the error frequency of the immersive content stream below one in a million to prevent noticeable frame loss, quality reduction, or visual artifacts that users might perceive.

\subsection{Future Network Congestion} As the number of users increase, the future wireless networks will face channel congestion and low-rank congested channel scenarios, where number of AP antennas are far less in number than active users. Current OMA-based methods used in industry standards do not achieve optimum data rates for low-rank congested scenarios and degrade in performance as the number of users increases. NOMA has been proposed for better bandwidth utilization, but fails to achieve high data rates for the low-rank-channel scenario. NOMA assumes a heuristic decoding order for each user, based on the absolute value of the channel coefficient scalar for that user. This does not hold when the channel is a vector for multiple antennas at the AP and/or each user.This network capacity degradation on rank-deficient channels is increasingly likely for future wireless networks. Furthermore, to achieve the same data rate with present systems, higher SNRs are required that degrade the bat

\section{Energy Efficient Solutions for Wireless Virtual Reality}
\subsection{Optimum GDFE Designs} 
Non-linear AP receivers are better for addressing congestion in low-rank channel scenarios, which are increasingly common in future wireless networks. GDFE exploits inter-user interference through successive interference cancellation, enabling significantly higher data rates at lower power levels. Wireless VR applications require high speeds (500+ Mbps) for seamless performance at low transmit-power levels. Generalized decision feedback equalizers (GDFE) outperforms linear receivers used in OMA for these scenarios, as in section\ref{sec: Caste study}'s algorithm, that optimizes power allocation and decoding orders. The GDFE’s ability to use all resource blocks and manage user interference yields significant benefits, including higher data rate sums compared to OMA, NOMA, and MC-NOMA, along with 5x energy savings for the same data rate, as shown in section \ref{sec: Caste study}. GDFEs use successive interference cancellation (SIC), decoding stronger signals first and subtracting them to minimize interference, unlike linear receivers that treat interference as noise, degrading performance. Achieving high sum rates at low power is vital for battery longevity in wireless VR, making GDFE-based power allocation methods crucial. These GDFE's help exploit the best user order and spatial energy allocation. A suboptimal decoding sequence can lead to interference that severely limits achievable throughput, especially when the channel is vector-valued and lacks a natural order—as is the case with multi-antenna APs. Therefore, determining the optimal decoding order becomes essential to fully exploit the spatial degrees of freedom and inter-user crosstalk.

\subsection{Multi-user encoding (MUE)} 
Wireless VR streaming demands high bandwidth and computational power. Multi-User Encoding (MUE) optimizes these resources by using correlations in users' views, reducing pixel requirements by 49\% compared to single-user encoding \cite{8038375}. This approach enables real-time collaborative VR by reducing data transmission and processing needs. MUE employs a hybrid unicast-multicast model, where a primary view is multicast to all users, and only view differences are unicast individually, cutting transmission time and conserving energy—crucial for battery-powered headsets. It also dynamically allocates resources based on users’ focus areas, enhancing compression when users view similar regions. As user numbers increase, MUE scales efficiently, maintaining consistent energy use per bit and improving collaborative VR support \cite{7539092} in distributed Wi-Fi scenarios.

\subsection{Edge/Cloud Computing} 
Wireless VR demands intensive processing for rendering and tracking, which can strain local devices, causing overheating and battery drain. Offloading these tasks to edge computing reduces processing time by 81\% and latency by 20-80\%~\cite{8038375, 9430902} compared to cloud computing, while compressing frames before transmission saves bandwidth by up to 35x. Edge servers, with motion-to-photon delays under 20 ms~\cite{8329628}, support seamless handover, shared VR spaces with latencies below 50 ms, and joint optimization across computing, caching, and communication layers for enhanced performance. Additionally, the caching of frequently accessed VR content at the edge minimizes retrieval times, reduces redundant transmissions, saves bandwidth, and ensures consistent frame rates and high-quality visuals. By storing stable components locally, caching speeds cloud-based rendering, balances server loads, and avoids network bottlenecks, meeting VR's ultra-low latency needs. 
Edge nodes can perform viewport-aware streaming by transmitting only the portion of 360$^\circ$ video within the user's current field of view, reducing bandwidth.  
They can also leverage specialized hardware like GPUs and VPUs optimized for rendering, tracking, and VR-specific workloads.  
This localized compute capability enhances responsiveness and visual quality while preserving battery life on lightweight VR devices.  
Together, these advantages make edge/fog computing essential for delivering scalable, high-fidelity immersive experiences in mobile environments.

\section{Case Study: Energy-Sum minimization for Low Rank Wi-Fi Channels}
\label{sec: Caste study}
\subsection{System Model}
\begin{figure*}[t!]
    \centering
    \includegraphics[width=1\linewidth]{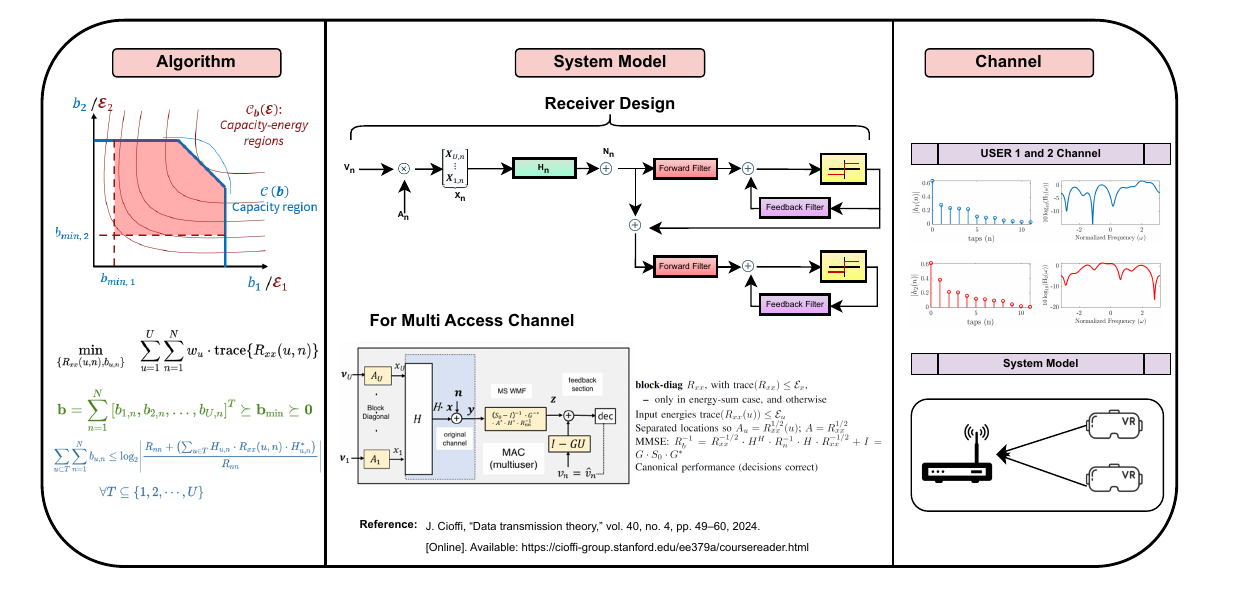}
    \caption{Complete System Model}
    \label{fig:hrl_structure}
\end{figure*}

The goal is to optimize the energy of AR/VR users while achieving the desired data rates by solving Fig. \ref{fig:hrl_structure}'s convex optimization problem. Various schemes have been proposed for this purpose, including OMA, NOMA, and MC NOMA. OMA, typical of today's most advanced Wi-Fi and cellular enterprise systems, allocates separate, non-overlapping resources, such as time, frequency, or code to each user to prevent interference, whereas NOMA allows users to share available resources. Although OMA simplifies resource allocation by isolating users, it can limit spectral efficiency and the overall data rate as the number of users grow. NOMA is only a power domain allocation, which assigns all subcarriers to all users involved, which results in suboptimal data rates. For MC-NOMA, the existing research uses heuristic SIC decoding order based on the ordering of the channel coefficient. While this may be optimal for SISO case, there is no natural order given by the channel vectors for each user in a MIMO scenario. This leads to usage of various heuristics for SIC order, e.g. norm of channel vector, etc, which are not optimal and lead to suboptimal data rates and high energy. The system model for this work's case study consists of 3 users, 2 AP antennas inside a room (simulation scenario IEEE 302.11b). The transmit power for the Wi-Fi AP is set at 17dBm.


\subsection{Proposed Framework}
NOMA allocates equal power across all subcarriers, resulting in suboptimal energy distribution per user. Current MC-NOMA research relies on heuristic SIC decoding orders based on the norm of channel vectors instead of deriving the optimal order, leading to inadequate data rates for upcoming wireless VR applications. This paper proposes and implements optimal power and subcarrier allocation for APs with multiple antennas. By eliminating heuristic decoding order assumptions and deriving the optimal order, the method decouples power-subcarrier allocation from decoding order, maintaining convexity and enabling efficient, optimal solutions.
In some cases, identical Lagrange multipliers for multiple users prevent individual decoding orders from achieving the required data rates~\cite{book}-~\cite{1494444}. To address this, we introduce time-sharing, which adaptively combines multiple decoding orders to achieve higher data rates than single-order NOMA methods. Our algorithm, minPMAC, achieves 39\%, 28\%, and 16\% higher data rates compared to OMA, NOMA, and MC-NOMA baselines under varying SNR levels, respectively.
Furthermore, section V proposes a near optimal, runtime efficient DRL-based algorithm for low rank channel congestion scenarios. The algorithm uses proximal policy optimization (PPO) as a DRL agent to maximize the energy efficiency. The state space considers transmit power and achieved data rates. The reward function maximizes data rate sum while minimizing transmit power. The PPO agent's clipped subjective surrogate objective function allows small steps to alter transmit powers as per the achieved data rates. DRL agents run faster inference on similar channel conditions after learning the required objective policy function. This near-optimal DRL formulation is named DRL-minPMAC.
\subsection{Simulation Setup}
The experiments use the WINNER\_A1\_LOS channel model with a carrier frequency of 5 GHz and a bandwidth of 80 MHz across 64 subcarriers. Both the AP and users have omnidirectional antennas. Channel realizations are generated using QuaDRiGa, an open-source library in MATLAB used for simulating radio channel models in wireless communication systems. Experiments generate 1,000 channel samples and average the results through Monte Carlo simulation. The noise power spectral density is -174dBm and the SNR range is from [-10, 50] dB.

\subsection{Results and Discussion} 
\begin{figure*}[t!]
     \centering
    \begin{subfigure}[t]{0.66\columnwidth} 
         \centering
         \includegraphics[width=\textwidth]{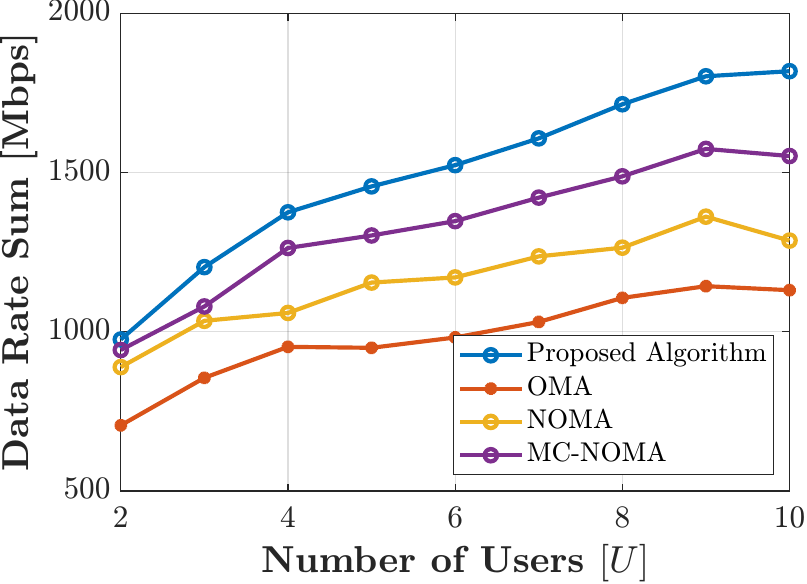}
         \caption{Number of users [$U$] vs. Sum rate [Mbps].}
         \label{fig:crosstalk}
     \end{subfigure}
     \hspace{-0.01\columnwidth} 
     \begin{subfigure}[t]{0.66\columnwidth} 
         \centering
         \includegraphics[width=\textwidth]{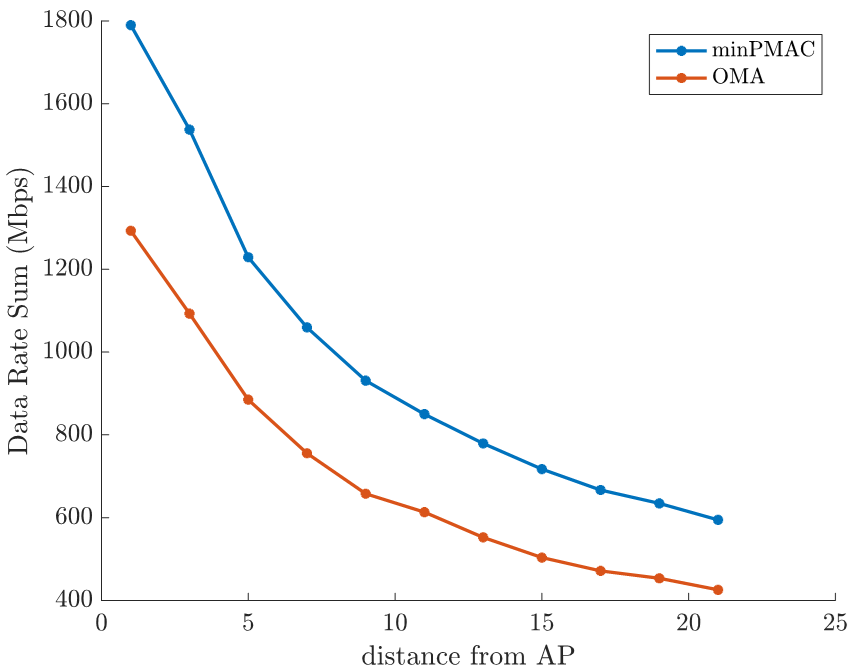}
         \caption{Distance from AP vs. Data Rate Sum}
         \label{fig:outage}
     \end{subfigure}
     \hspace{-0.01\columnwidth} 
     \begin{subfigure}[t]{0.66\columnwidth} 
         \centering
         \includegraphics[width=\textwidth]{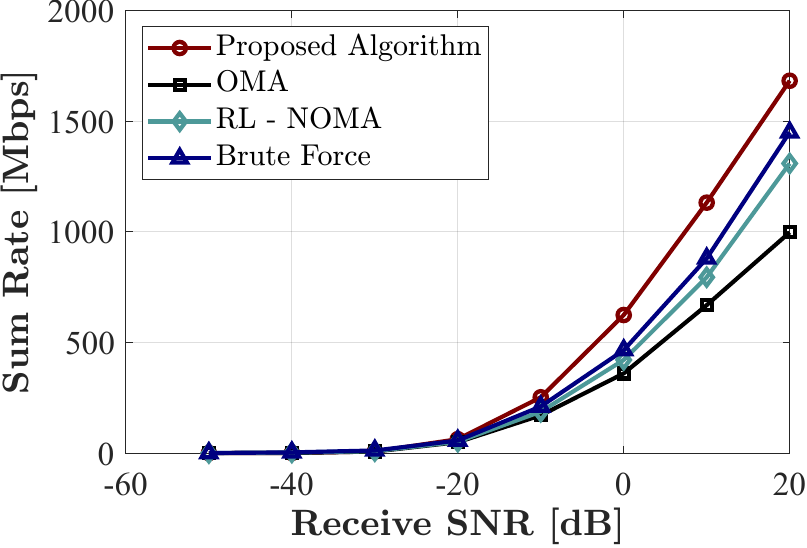} 
         \caption{Recieve SNR [db] vs. Sum rate [Mbps]}
         \label{fig:Sumrate}
     \end{subfigure}
      \caption{\textbf{(a)} Number of users vs. Data rate sum [Mbps] for users at distance 3m from the AP compared to baseline methods [OMA, NOMA and MC-NOMA], \textbf{(b)} Distance from AP vs. Data Rate Sum (Mbps) with path loss exponent = 4,\textbf{(c)} Receive SNR [db] vs. Sum rate [Mbps] for 3 users located at distances [3m, 3m, 3m] from the AP compared to baseline DRL methods [Brute Force, DRL-NOMA and OMA]}
     \label{fig:2}
\end{figure*}

Fig.~\ref{fig:crosstalk} shows the effect of cross talk as the number of users increase. OMA, which is the currently implemented industry standard, deteriorates in performance as interference increases. NOMA and MC NOMA, as used in current research, also fail to achieve optimum data rates because they do not incorporate optimum decoding order, but rather depend on heuristic decoding order assumptions. The proposed algorithm demonstrates highest rate sum because of optimum decoding order and time-sharing. The ability of the non-linear processing to take advantage of increasing cross talk as users increases helps mitigate the effect of low-rank channels. This is extremely useful and practically applicable for collaborative VR environments.

Fig.~\ref{fig:outage} describes the distance from AP vs. Data rate sum with the path loss exponent 4. We observe that minPMAC demonstrates 10x energy savings as compared to current cellular standard (OMA). This comparison demonstrates the effectiveness of the proposed resource allocation approach in achieving the minimum transmission energy under the required data rate constraints.

Fig.~\ref{fig:Sumrate} compares the data sum-rate comparison of the minPMAC with the DRL-minPMAC and with brute force. MinPMAC consistently outperforms DRL-minPMAC uplink NOMA and brute-force. The baseline method used to compare to DRL-minPMAC is a brute force algorithm that conducts a heuristic search over the power-level allocation space. MinPMAC  algorithm, due to its optimal decoding order derivation for vector channel, as well as time-sharing, achieves optimum rate sum for all SNRs as per AR/VR requirements.

MinPMAC contributes significantly to the AR/VR requirements discussed in Section II and the challenges discussed in Section III. Firstly, minPMAC and DRL-minPMAC optimize energy levels and address the power constraints while maximizing data rates to achieve the required FOV, resolution, and refresh rate. DRL-minPMAC effectively addresses convergence issues, achieving a 5x speedup over the optimal minPMAC, significantly reducing latency. While this improvement is substantial, ideal resource allocation would occur within the channel's coherence time to fully align with real-time constraints.

\section{Future 6G Research Directions and Challenges for Metaverse}
\label{sec: Virtual reality}
This section explores the key technologies that enable the Metaverse, particularly focusing on the challenges and open research problems to realize wireless VR through next-generation Wi-Fi and cellular enterprise networks.
\subsection{Integrated Sensing and Communications (ISAC)}

ISAC integrates sensing capabilities with communication functions within a unified framework, essential for the precision and low-latency requirements of wireless VR. It aims to develop sensing technologies that monitor the physical world using 6G communications. By sharing wireless spectrum and hardware resources, ISAC enables efficient data collection and real-time feedback, crucial for maintaining high-quality virtual experiences. For high-level interactivity, such as training simulations and gaming, the user experience relies on low-latency, accurate feedback made possible by mapping physical movements and environmental changes into the virtual space through ISAC~\cite{electronics12173651}. Additionally, next generation Wi-Fi networks seek to integrate communications, sensing, computing, and storage into a unified system, providing the infrastructure necessary for the metaverse and wireless VR to thrive. However, ISAC faces several challenges in making wireless VR a reality, including developing systems that achieve precision without compromising communication quality, creating energy-efficient architectures to support high computational demands, and ensuring data privacy against potential threats.


\subsection{Reconfigurable Intelligent Surfaces (RIS)}
Reconfigurable intelligent surfaces (RIS) are transforming next-generation wireless networks by enhancing reliability, latency, and coverage capacity. RISs establish virtual line-of-sight (VLOS) connections between users and base stations in challenging environments by adjusting the phase of incoming signals through their electromagnetic properties, enabling passive beamforming toward target users. When line-of-sight (LoS) paths are blocked, RIS maintains consistent data rates essential for applications like video streaming at 30 frames per second, requiring local display rates of 30 Gbps~\cite{9906483}. Supporting such high data demands across millions of devices necessitates substantial spectrum resources. However, integrating RIS with wireless VR presents challenges, including high mobility that complicates channel state information (CSI) acquisition and affects beamforming precision, real-time processing for hundreds of antennas leading to scalability and interference issues, rapid RIS adjustments for user tracking, optimal RIS placement for comprehensive VR coverage, and potential increases in energy consumption for large deployments.

\subsection{Semantic Communications} 
Semantic communications is an emerging field that transcends traditional entropy-based compression by conveying the essential meaning or intent of information instead of raw data. This user-centric approach focuses on transmitting core concepts and contextual significance, making it ideal for high-data-rate, bandwidth-intensive applications like wireless VR. It optimizes tasks such as 3D mesh recovery for human bodies and 3D scene reconstruction for virtual object placement~\cite{Xia_2023}. Key strategies include perceptual quality optimization (e.g., LPIPS, GAN-based losses) to prioritize visually salient VR regions, adaptive coding schemes for efficient resource use, and view synthesis to address field-of-view mismatches, thereby reducing latency and enabling local adjustments. However, challenges persist, including accurately extracting task-relevant semantic information, developing metrics to assess transmitted data quality, ensuring resilience against distortions, and leveraging AI for semantic encoding, which demands advanced algorithms and significant computational power.

\section{Conclusion}
\label{sec:conclusion}
The pursuit of truly immersive wireless VR experiences necessitates a radical departure from traditional wireless communication paradigms. While advancements in areas such as RIS, ISAC, and semantic communications offer promising avenues, our analysis reveals that a fundamental shift in receiver design, specifically through the adoption of optimized nonlinear techniques like GDFE, is crucial for overcoming the inherent limitations of low-rank channels, especially in congested environments. The practical gains, demonstrated in our case study with the minPMAC algorithm’s superior performance and the speed of the DRL-minPMAC approach in Wi-Fi scenarios, highlight the potential of this approach, moving us closer to realizing the high data rates and low energy consumption that are paramount for widespread wireless VR adoption. Future research must focus on refining these nonlinear techniques for real-time operation, especially within the strict latency bounds dictated by human perception. The ongoing development of resource-allocation strategies that are not only optimized for individual user experiences but also for large scale collaborative VR settings will be the driving force of true metaverse realization.

\section{Acknowledgements}
This research is supported by Stanford in collaboration with Intel Corporation, Samsung Semiconductors and Ericsson.

\bibliographystyle{ieeetr}

\vskip -2\baselineskip plus -1fil
\begin{IEEEbiographynophoto}{Muhammad Ahmed Mohsin} (\href{mailto:muahmed@stanford.edu}{muahmed@stanford.edu}) received the B.E. degree in electrical engineering from National University of Sciences and Technology (NUST), Pakistan in 2024. He is currently pursuing the Ph.D. degree in electrical engineering from Stanford University, USA. His primary focus of research lies in next-generation wireless communications, 3D computer vision for perception and deep reinforcement learning.
\end{IEEEbiographynophoto}

\vskip -2\baselineskip plus -1fil
\begin{IEEEbiographynophoto}{Sagnik Bhattacharya} (\href{mailto:sagnikb@stanford.edu}{sagnikb@stanford.edu}) received the B.Tech degree in Electrical Engineering from Indian Institute of Technology, Kanpur, India, in 2021. He worked at Samsung Research, Seoul, South Korea, from 2021-2023 as a Wireless Research Engineer. He is currently pursuing a Ph.D. in Electrical Engineering from Stanford University, USA. His primary research interests are machine learning-enabled communications, wireless system design, information theoretic generative AI, and joint source channel coding.
\end{IEEEbiographynophoto}

\vskip -2\baselineskip plus -1fil
\begin{IEEEbiographynophoto}{Abhiram R. Gorle} (\href{mailto:abhiramg@stanford.edu}{abhiramg@stanford.edu}) earned his B.Tech degree in Electrical Engineering from the Indian Institute of Technology, Madras, India, in 2024. He is currently pursuing an M.S. degree in Electrical Engineering at Stanford University, USA. His research interests include Information Theory, Wireless Communications, Signal Processing, and their interplay with foundational machine learning models.
\end{IEEEbiographynophoto} 


\vskip -2\baselineskip plus -1fil
\begin{IEEEbiographynophoto}{Muhammad Ali Jameshed} (\href{mailto:muhammadali.jamshed@glasgow.ac.uk}{muhammadali.jamshed@glasgow.ac.uk}) (Senior Member IEEE) received a Ph.D. degree from the University of Surrey, Guildford, U.K, in 2021. He is with University of Glasgow, since 2021. He is endorsed by Royal Academy of Engineering under exceptional talent category and was nominated for Departmental Prize for Excellence in Research in 2019 and 2020 at the University of Surrey. He is a Fellow of the Royal Society of Arts, a Fellow of the Higher Education Academy UK, a Member of the EURASIP Academy, and an Editor of the IEEE Wireless Communication Letters and an Associate Editor of the IEEE Sensor Journal, IEEE IoT Magazine, and IEEE Communication Standards Magazine. His research interests are energy efficient IoT networks, AI for wireless communication, EMF exposure measurements, and backscatter communications.
\end{IEEEbiographynophoto} 

\vskip -2\baselineskip plus -1fil
\begin{IEEEbiographynophoto}{John M. Cioffi} (\href{mailto:ali.cioffi@stanford.edu}{cioffi@stanford.edu}) (Life Fellow, IEEE) has received his B.S. in electrical engineering from the University of Illinois in 1978 and his Ph.D. from Stanford in 1984. He worked at Bell Labs and IBM Research before founding Amati Communications in 1991 (later acquired by TI). A Professor Emeritus at Stanford, he is also the Chair and CEO of ASSIA Inc. With over 800 publications and 150 licensed patents, Cioffi has received numerous honors, including the IEEE AG Bell Medal, the 2023 US National Medal of Technology from the USA's President, and the Marconi Fellowship. He currently chairs PhyTunes and the Marconi Society.
\end{IEEEbiographynophoto}
\bibliography{references/ref.bib}

\end{document}